\documentstyle[aps,prl,epsf]{revtex}

\newcommand{\ba}{\begin{eqnarray}}
\newcommand{\ea}{\end{eqnarray}}
\newcommand{\be}{\begin{equation}}
\newcommand{\ee}{\end{equation}}

\begin{document}

\title{Theory of the Breakdown of the Quantum Hall Effect}
\author{V. Tsemekhman, K. Tsemekhman, C. Wexler, J.H. Han, and D.J. Thouless}
\address{Department of Physics, University of Washington, 
	Box 351560,Seattle, WA 98195}

\date{June 1996}

\maketitle

\begin{abstract}
Breakdown of the Quantum Hall Effect at high values of injected
current is explained as a consequence of an abrupt formation of
a metallic ``river''
percolating from one edge of the sample to the other. Such river is
formed when lakes of compressible liquid, where the long-range disorder
potential is screened, get connected with each other due to the strong
electric field. Our theory predicts critical currents consistent with
experiment values and explains various features of the breakdown.
\end{abstract}

\section*{}
Since the discovery of the Quantum Hall Effect (QHE)~\cite{klit1}
great variety of phenomena associated with the behavior of the
two-dimensional electron gas (2DEG) in the strong magnetic field has
been studied both experimentally and theoretically. In this paper we
discuss the breakdown of the QHE at high injected currents.  First
observed by Ebert~\cite{bd} as a sudden onset of the dissipation when the
injected current exceeds some critical value, the
breakdown of the integer QHE (IQHE) has been intensively investigated since
then~\cite{111}. While the absence of the  
dissipation ($\sigma_{xx}=0$) and the quantization of the transverse
conductivity $\sigma_{xy}$ in the IQHE regime at low
temperatures are well understood, the breakdown of the dissipationless
regime has not been given a clear explanation. Existing theories of
this effect~\cite{qhe} include production
of hot electrons~\cite{hot-e}, inter-Landau level transitions~\cite{ll-tr}
in the high local electric field (due to tunnelling or emission of
phonons), increase in the number of the delocalized states
in the Landau level~\cite{del-st}. None of these produced results
consistent with experiments: except for the hot-electron
picture, all others predict values of the critical current density
that are two orders of magnitude
higher than those observed experimentally. Also, neither of them
can explain hysteresis and localized nature of the breakdown
phenomenon, nor the existence of the
transient switching and broadband noise before the breakdown, as well
as peculiar steps in the magnetic field dependence in the regime of
critical current. Besides, all
these theories need to use some artificial assumptions and are too
complicated to be correct.  

Experiments on the breakdown of the QHE are performed on GaAs
heterostructures or on high-quality MOSFET devices both characterized
by the presence of long-range fluctuations of the disorder
potential. In the GaAs systems the disorder potential due to the remote
dopants has predominantly long wave-length
fluctuations ($\lambda>d>l_{H}$, where $\lambda$ is the wavelength of
the fluctuations, $d$ is the spacer thickness, and $l_{H}$ is the
magnetic length). Long-range potential fluctuations are also present
in high-mobility MOSFET devices though their nature is not that
evident~\cite{vK84}. When the magnetic field and the density of
the 2DEG correspond to a filling factor close to an integer
number, these fluctuations are not screened by the electrons in 
most parts of the system~\cite{efros}. The region with completely
filled Landau level percolates through the sample, leading to the
QHE (at the end of this paper we discuss what happens
when the filling factor is far from an integer). In the percolating
incompressible region electric fields are typically about the
same as those of the bare disorder potential~\cite{chalker}
$E_{inc}\simeq\sqrt{<{E_{bare}^2}>}=\sqrt {n_0/32
\pi} e/\epsilon \epsilon_0 d$. For a typical clean
sample this leads to: $E_{inc}\simeq
0.1\hbar\omega_{c}/el_{H}$, where $\omega_{c}$ is the cyclotron frequency.
There are, however,
isolated areas where the long wave-length fluctuations of the disorder
potential are screened. The temperatures in the experiments on the
breakdown ($T \sim 1$ K) are high compared to the energy scales of both
the shorter wave-length fluctuations (that are left unscreened) and of 
the residual inter-electron interaction. Under such
conditions a compressible liquid fills these isolated areas of
screened potential which, therefore, behave like
metallic lakes. In each of those lakes, all states of the highest
available Landau level are partially occupied, and the screened
potential fluctuates around the Fermi level $\epsilon_{F}$ with an amplitude
of the order of $T$ which is much smaller than the amplitude of the
potential fluctuations in the incompressible region. Below we consider
the potential in the compressible lakes to be flat. The main idea of our
theory is that at high 
enough currents the insulating region separating two lakes, suddenly
breaks down due to the high electric field; this leads to the connection
of the lakes. When such connected lakes form a metallic river
percolating from one edge of the sample to the other, abrupt onset of
the dissipative regime is observed.

In general, the distance between two adjacent lakes can be large,
and there can be several fluctuations of the potential in the
incompressible region separating two lakes. We will see, however, from
the results of the numerical simulations
that the lakes closest to each other get connected first. Let us,
therefore, consider a simple model in which two metallic lakes are
separated by a narrow insulating region with one parabolic potential
fluctuation characterized by the root mean square electric field
$E=E_{inc}\simeq 
0.1\hbar\omega_{c}/l_{H}$. In equilibrium, all electrons in the
lakes have energies equal to $\epsilon_{F}$ (up to $T$), and all 
electrons in the incompressible region  separating them  have energies
below the Fermi energy. Electrons
cannot move from the lake into the incompressible region because
no states are available there (except on the next Landau level, which
is much higher in energy). They do not flow into the lake either,
because the energy of any available state $\epsilon_{F}$ is higher than that
of any state in the completely filled region. Injection of the current
into the 2DEG leads to the appearance of the Hall electric field
troughout the sample. Assume that the direction of the electric field
coincides with the direction from one lake to the other. In the
insulating region this external
field can be considered to be uniform on the scale of the distance
between the lakes. In the lakes the charges are redistributed to
screen the external field.  Now, however, the two lakes are at
different potentials (Fig. 1a). One can easily show
that when the external field $E_0=\sqrt{3}E_{inc}$, the minimum of the
parabolic potential fluctuation shifts to the boundary of the metallic
lake, so that the potential monotonically drops from one lake
towards the other.  At this point, electrons can freely
move from the incompressible region into the lake, as their energies
are higher than those of the partially filled states in the lake. In
this simplified picture, $E_0$ is a critical field at which the two
lakes get connected. Possible existence of several
fluctuations in the separating region does not change the
picture. Lakes get connected when the potential drops monotonically
between them, i.e. when the last (the fastest) fluctuation is smoothed out
by the electric field. The estimate for the critical field remains
basically the same because the probability for the electric field of
the disorder potential
to be larger than $0.2\hbar\omega_{c}/l_{H}$ is exponentially
small.

The above value of the electric field necessary to connect the
lakes does not need to be equal to
the critical average electric field $E_{c}=V_{H}/W$, where $V_{H}$ is
the measured Hall voltage, and $W$ 
is the width of the sample. It is easy to see that the electric field
in the narrow region between 
the two lakes can be much larger than the average Hall electric
field: if the size of the separating region is
smaller than the size of the lakes, the equipotential lines are highly
squeezed between the two lakes. Therefore, an average electric field
{\em much smaller} than $E_0$ is needed to connect the lakes.  When the two
lakes merge, the electric field is expelled from the former
incompressible region now covered with compressible liquid. Therefore, the
electric field becomes stronger outside the newly formed large lake,
facilitating further connections as the injected current is increased.
One would expect that 
after the metallic region acquires some critical length in the
direction of the Hall electric field upon the increase of the injected
current, the process of
further connection must take an avalanche form. Finally, when the
compressible liquid forms a river flowing from one edge to the other,
dissipation drastically increases. It is important to notice that 
just before the formation of the metallic river, the longitudinal
conductivity is still determined by the variable-range hopping; also,
the effective size of the insulating system is still
quite large at this moment,
as there are many lakes on the way of the future river that are not
yet connected. This is the reason why the longitudinal conductivity
jumps several orders of magnitude at the critical current: the
mechanism of the conductivity abruptly switches from hopping to
the metallic one.

\begin{figure}
	\begin{center}
	\leavevmode
	\epsfbox {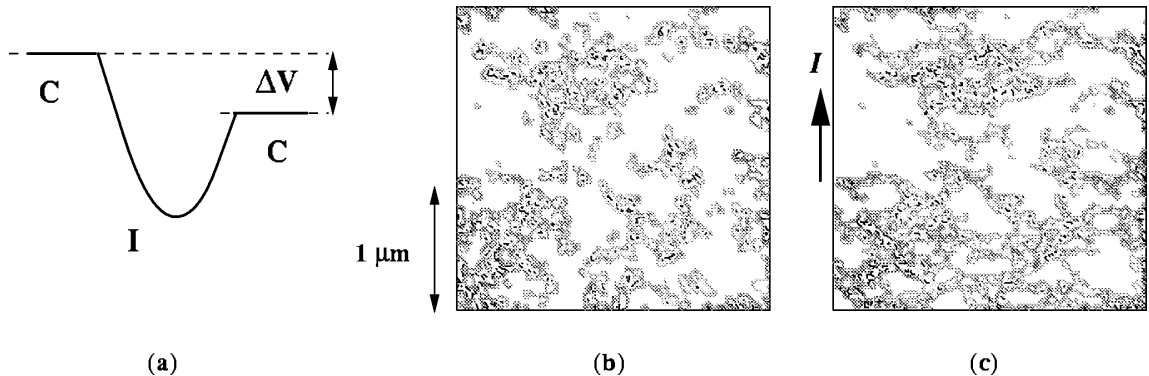}
	\end{center}
\caption {a. Distribution of the potential in the two adjacent islands
and between them before breakdown; b. and c. Charge distribution from
numerical simulations for the $2.5$ $\mu$m $\times 2.5$ $\mu$m sample with 
$d=25$ nm, $l_{H}=10$ nm, $\nu=1.93$, $T=0.5$ K. 
White areas are the $\nu=2$ incompressible liquid, grey
areas the $\nu<2$ partially filled compressible liquid. b. at zero
electric field, c. at $E=500$ V/cm} 
\end{figure}   

To see how the breakdown occurs in real systems we performed
numerical simulations (the details will be published elsewhere). 
In equilibrium, the free energy of the interacting 2DEG 
in the strong magnetic field and in the smooth disorder potential was
minimized with respect to the occupation numbers at finite
temperature. The disorder potential was created by a randomly chosen
distribution of charged impurities in a plane a distance $d$ away
from the 2DEG. Coulomb interaction between electrons 
was taken into account in the mean-field approximation. Fig. 1b shows
the charge distribution for a clean sample at $\nu=1.93$. Lakes of 
compressible liquid ($\nu<2$, grey areas) are immersed into the percolating
incompressible region ($\nu=2$, white areas). The effect of the Hall
current shows up in the appearance of the external electric field
which is taken to be uniform before the response of the charges in the
system. Relaxation current is allowed to flow only in the compressible
regions. The charge density of the same system but with external field of
$500$ V/cm is shown on Fig. 1c. Evidently, this field is already higher
than the critical one: a metallic river is well established. This
is in good agreement with experimental data. 

The fact that breakdown is localized in space directly follows from
our theory: a metallic river is very narrow at the 
breakdown. Connection of the lakes at fields below the critical one
causes charge redistribution and
energy relaxation. The amount of dissipated energy varies with the
size of the connecting lakes. This energy relaxation shows up as a broadband 
noise. The transient switching is the result of the formation and
immediate disconnection of the river at the fields just below
critical. This happens since after the river is formed, relaxation is
available all over the sample width. The system can then relax to a
new steady state with incompressible region still
percolating. Our numerical simulations show such connection/disconnection
processes. As in
any dielectric breakdown, hysteresis is the result of the irreversible
change of the system properties (namely the pattern of rivers and
lakes) after the current was allowed to flow through the system. 

Finally, the peculiar steps observed in the magnetic field dependence of
the longitudinal voltage in the critical current regime show the 
evidence of the opening of new metallic channels when the system
is moved away from the center of the plateau. In equilibrium, as the
difference of the
filling factor from the integer becomes larger, the size of the lakes
of compressible liquid increases while their separation becomes
smaller. Therefore, in the critical
regime the number of the percolating narrow metallic channels also
grows, each giving its own discrete contribution in the voltage
drop. The question of what happens if at certain filling factor
compressible region percolates even in equilibrium, will be addressed
in another paper. Here, however, we want to mention that
even at zero temperature in
the mean-field approximation and with only long-range disorder, the
relative width of the QH plateau is less than $0.4$, and can be very
small for clean samples. While the residual electron-electron interaction
leads to the fractional QHE (FQHE) in very clean samples, the systems not
exhibiting FQHE will still show step-like transition between QH
plateaux. Such behavior can be observed only at very low temeperatures
when the fluctuations of the screened long-range potential in the
compressible region ($\simeq T$) become comparable with the amplitude
of {\it the short wave-length ($<l_{H}$)} disorder potential fluctuations
which, though being exponentially small, remain unscreened. At such
low temperatures the properties of the system are determined by the
short-range random potential.      

In conclusion, we propose a theory of the breakdown of the QHE based
on the existence of the compressible regions in the inhomogeneous
2DEG. The predictions of the theory agree with experimental data. 

This work was supported by the NSF grant DMR-9528345.


\references

\bibitem{klit1} K. von Klitzing, G. Dorda, and M. Pepper,
Phys. Rev. Lett. {\bf 45}, 449 (1980). 
\bibitem{bd} G. Ebert {\it et al}, J. Phys. C {\bf 16}, 5441 (1983).
\bibitem{111} M. Cage, J. Res. Natl. Inst. Stand. Technol. {\bf 98},
361 (1993), and references therein.
\bibitem{qhe} The Quantum Hall Effect, ed. by R. E. Prange and
S. M. Girvin, Springer-Verlag, New York (1990), and references therein. 
\bibitem{hot-e} S. Komiyama, T. Takamasu, S. Hiyamizu, S. Sasa,
Solid State Comm. {\bf 54}, 479 (1985); Surf. Sci. {\bf 170}, 193 (1986). 
\bibitem{ll-tr} D.C. Tsui, F. Dolan, A.C. Gossard,
Bull. Am. Phys. Soc.
{\bf 28}, 365 (1983); O. Heinonen, P.L. Taylor, S.M. Girvin,
Phys. Rev.  {\bf 30}, 3016 (1984).
\bibitem{del-st} S. Trugman, Phys. Rev. B {\bf 27}, 7539 (1983).
\bibitem{vK84} K. von Klitzing {\it et al}, Proc. of the 17th Int. Conf. on
the Physics of Semiconductors, ed. D. J. Chadi and W. A. Harrison, 1984.
\bibitem{efros} A. Efros, Solid State Commun., {\bf 65}, 1281 (1988); 
A. Efros, Solid State Commun., {\bf 67}, 1019 (1988). 
\bibitem{chalker} N. R. Cooper and J. T. Chalker, Phys. Rev. B, {\bf
48}, 4530 (1993).
\end{document}